\documentclass[aps,twocolumn,showpacs,showkeys,amsmath,amssymb,superscriptaddress,nofootinbib,floatfix,longbibliography,noeprint
]{revtex4-2}
\usepackage{amsmath,graphicx,color,ulem}
\usepackage{epsfig,bm,slashed,hyperref}

\DeclareMathOperator{\Tr}{Tr}

\begin{document}

\title{Magnetic susceptibility of a hot hadronic medium and quark degrees of freedom \\ near the QCD cross-over point}

\author{Rupam Samanta}
\email{rupam.samanta@ifj.edu.pl}
\affiliation{Institute of Nuclear Physics, Polish Academy of Sciences,  31-342 Cracow, Poland}

\author{Wojciech Broniowski}
\email{wojciech.broniowski@ifj.edu.pl}
\affiliation{Institute of Nuclear Physics, Polish Academy of Sciences,  31-342 Cracow, Poland}

\begin{abstract}
The lattice QCD results for the temperature-dependent magnetic susceptibility of the medium below the cross-over temperature 
are not possible to reconcile with the widely used Hadron Resonance Gas model, also amended with the physical magnetic moments of hadrons 
or the pion--vector-meson loops. As noticed earlier, one observes a substantially too strong diamagnetism 
at temperatures in the range above $\approx 120$~MeV compared to the lattice. 
This hints at a presence of quarks significantly below the QCD cross-over temperature, which are needed as a source of paramagnetism. However,  
the pions must be retained to describe the diamagnetism data at low temperatures.
Therefore, we consider here a quark-meson approach, where the temperature-dependent quark masses are fixed in a model-free way using the
baryon-baryon and baryon-strangeness susceptibilities from the lattice at zero magnetic field. The constituent quarks possess anomalous magnetic moments estimated from 
the octet baryon magnetic moments. 
The vacuum quark-loop and meson-loop contributions are duly incorporated. We show that in such a framework,
one can describe the magnetic susceptibility up to the cross-over point. The qualitative conclusion is that 
the QCD degrees of freedom must extend far below the cross-over temperature, down to $\approx 120$~MeV.
\end{abstract}


\maketitle

\section{Introduction\label{s:intro}}

Magnetic properties of the hot strongly interacting medium
have been actively studied on the lattice~\cite{Bali:2011qj,Bali:2012jv,DElia:2012ems,Endrodi:2013cs,Bonati:2013lca,Bonati:2013vba,Bali:2014kia,Bali:2020bcn,Endrodi:2022wym,Ding:2023bft,Brandt:2024blb,Endrodi:2024cqn,Astrakhantsev:2024mat,Ding:2025jfz}; for recent reviews, see~\cite{Andersen:2014xxa,Adhikari:2024bfa}.
Apart from their fundamental nature, they are associated with very interesting physical phenomena, 
such as the chiral magnetic effect~\cite{Kharzeev:2007jp,Fukushima:2008xe,Kharzeev:2015znc}, or the 
magnetic catalysis and inverse catalysis (for a recent review, see, e.g.,~\cite{Andersen:2021lnk}), reflecting the dependence 
of the quark condensates on the magnetic field. 
The information from the lattice QCD at non-zero temperature $T$, vanishing chemical potentials,
and non-zero magnetic field $B$, involves also properties such as the magnetic field dependent susceptibilities related 
to various conserved charges (baryon, strangeness, electric)~\cite{Ding:2023bft,Ding:2025jfz}, the chiral susceptibilities~\cite{Karsch:1994hm,Sasaki:2006ww,Bazavov:2011nk}, or magnetization~\cite{Bali:2014kia,Adhikari:2021bou}. A very 
basic and relevant quantity is the magnetic susceptibility, 
$\chi_B$~\cite{Bali:2012jv,Endrodi:2013cs,Bonati:2013vba,Bali:2020bcn,Endrodi:2022wym,Brandt:2024blb}, central to this work, 
as it describes the response of the medium to the magnetic field near $B=0$.

It is well known that at vanishing chemical potentials, QCD exhibits a cross-over 
phase transition~\cite{Stephanov:1998dy,Fodor:2004nz,Aoki:2006we,Fukushima:2010bq} near the
pseudo-critical temperature $T_c\approx155$~MeV, passing gradually from the hadronic degrees of freedom below $T_c$ 
to quarks and gluons above. The hadronic side is frequently described with the Hadron 
Resonance Gas model (HRG)~\cite{Fermi:1950jd,Pomeranchuk:1951ey,Hagedorn:1965st,Koch:1986ud}, 
involving stable hadrons and resonances, mimicking the inter-hadron interactions~\cite{Dashen:1969ep,Venugopalan:1992hy}.
HRG reproduces very well many features of the hot medium measured on the lattice~\cite{Karsch:2003zq,Gyulassy:2004zy}. In 
the absence of the magnetic field, the list includes the entropy density, the energy density and pressure, the sound velocity~\cite{Huovinen:2009yb,HotQCD:2014kol,Ding:2015ona}, the susceptibilities associated to conserved currents 
and their correlations, or the cumulants~\cite{Borsanyi:2011sw,HotQCD:2012fhj,Bellwied:2015lba,Borsanyi:2018grb,Bollweg:2021vqf}. 
At  $0<B\lesssim 0.15~{\rm GeV}^2$ (we use natural units of $\hslash=c=k_B=e=1$ throughout this paper),  
the conserved charge susceptibilities~\cite{Ding:2021cwv,Ding:2023bft,Ding:2025jfz} for $T\approx T_c$
are properly reproduced in HRG after the inclusion of the physical magnetic moments of hadrons~\cite{Samanta:2025mrq}.

With the above successes, HRG became a standard benchmark for the evaluation of thermodynamic quantities in the hadronic phase. 
Moreover, this is to be linked with the success in the description of the particle yields (see~\cite{Andronic:2017pug} and references therein) 
in ultra-relativistic nuclear 
collisions, where statistical hadronization with the chemical freeze-out
at $T \approx T_c$ leads to a remarkable agreement of the model with the experimental data. 
Consequently, a conviction followed that below $T_c$ one simply deals with just the hadrons, and above with the quarks and gluons. Certainly, the cross-over
nature of the transition makes the division smooth, with the hadrons and partons overlapping near $T_c$ within some range, 
but a practice developed that HRG can be applied up to $T_c$ with a stunning accuracy. 

However, there is a notable difficulty in the application of HRG 
to describe the lattice data for $\chi_B$ in the region below $T_c$, where it works stunningly well for other observables.
A comparison study was originally done in the seminal lattice papers~\cite{Bali:2014kia,Bali:2020bcn}, and more
recently in~\cite{Brandt:2024blb}. In these works, diamagnetism of the QCD medium was found at $T$ ranging from $\approx110$~MeV (lowest covered range) 
up to $\approx150$~MeV (the cross-over temperature), fulfilling the expectations based on the fact that the pion diamagnetism dominates at low temperatures.
Notably, it was a considerable qualitative improvement compared to earlier lattice QCD studies (cf.~\cite{Bonati:2013vba} and 
the comparison in Fig.~9 of~\cite{Bali:2014kia}), where $\chi_B$ was not displaying diamagnetism in the covered temperature range at all. 
Despite the success of obtaining proper agreement between the lattice QCD and HRG at temperatures $\approx 110$~MeV (where it is dominated by pions and kaons), 
the vivid mismatch at temperatures above $\approx120$~MeV proves that HRG is not adequate there.

In this paper, we first repeat the HRG study, showing that the mismatch persists with a complete list of hadronic 
states included and with physical magnetic moments of hadrons 
incorporated as in~\cite{Samanta:2025mrq}. 
In a novel study, we analyze the contribution to $\chi_B$ in HRG from pion--vector-meson loops, where
the resulting paramagnetic effects are small~(Sec.~\ref{s:loop}), at the level of 10-15\%, and agree qualitatively with the two-loop Chiral Perturbation 
Theory~\cite{Hofmann:2021bac}.  We also point out that the possible role of the hadron magnetic polarizabilities to $\chi_B$ 
is tiny~(Sec.~\ref{s:HRG_magpol}).

We then argue, on general grounds, that to be able to describe the lattice data for $\chi_B$ above $T\approx120$~MeV, a suitable model
needs to possess sufficiently {\it light paramagnetic} degrees of freedom, 
not to be damped thermally. Natural candidates here are the quarks of constituent masses of a few hundred MeV, 
appearing in the thermal system from the melt-down of baryons~(see, e.g., \cite{McLerran:2008ua,McLerran:2018hbz,Marczenko:2022jhl}). 
However, the pions must be kept, to retain the diamagnetism at lower temperatures. Hence, we apply the quark-meson 
framework (see~\cite{Schaefer:2004en} and references therein), which provides the needed paramagnetism in the vicinity of $T_c$, while pions and kaons supply diamagnetism at lower $T$. 
 
The methodology applied in our analysis is as follows: we use a non-interacting quark-meson model below $T_c$ 
to obtain $\chi_B$, while demanding a simultaneous description of the lattice data for the baryon-baryon ($\chi_{\cal{BB}}$) and 
baryon-strangeness ($\chi_{\cal{BS}}$) susceptibilities.\footnote{To our knowledge, no previous quark model calculations have been able to simultaneously describe
the lattice data for $\chi_B$, $\chi_{\cal{BB}}$, and $\chi_{\cal{B}S}$.} 
Specifically, we use the lattice data for $\chi_{\cal{BB}}$ and $\chi_{\cal{BS}}$ (at $B=0$)~\cite{Bollweg:2021vqf} to find the temperature dependence of the effective 
quark masses. Since these lattice studies are for $T>135$~MeV, we are able to fix the masses above this temperature. 
For $\chi_{\cal{BB}}$, and $\chi_{\cal{B}S}$, as we discuss in detail in Sec.~\ref{s:selfcons}, in the independent particle approach only the thermal 
part of the free energy contributes, because the vacuum part does not explicitly depend on the chemical potentials. This fact allows fixing the quark masses 
in a straightforward way from the thermal part of the free energy.  

With thus obtained temperature-dependent quark masses, and with the pions and kaons present, we compute the subtracted 
magnetic susceptibility $\chi_B(T)-\chi_B(0)$, as provided by the lattice simulations~\cite{Bali:2020bcn,Brandt:2024blb}.
The vacuum contributions, which here are relevant~\cite{Endrodi:2013cs,Kamikado:2014bua}, are incorporated. For the quarks, the
vacuum part requires a regulator in the spirit of the low-energy effective non-perturbative models~\cite{RuizArriola:2002wr,Broniowski:2021awb}, 
which eliminates the contribution of the hard modes. We also investigate the relevant role of the anomalous magnetic 
moments of the constituent quarks. 

The principal conclusion of our study is that the quark-meson model, with the vacuum terms included, complies with the lattice data for $\chi_B$ and  
simultaneously with $\chi_{\cal{BB}}$ and $\chi_{\cal{BS}}$ -- the task that has not been accomplished within HRG.

The article is organized in the following way: First, in Sec.~\ref{s:susc_res}, we discuss the lattice results 
for the conserved charge susceptibilities, $\chi_{\cal{BB}}$ and $\chi_{\cal{BS}}$, as well as
for the magnetic susceptibility, $\chi_B$, comparing them to the state-of-the-art HRG calculation and pointing out the above-discussed discrepancy in the case of $\chi_B$. 
We list the basic definitions of the thermodynamical framework in Sec.~\ref{s:theory}. In Sec.~\ref{s:HRG}, we discuss the HRG model and list the
formulas for the susceptibilities in the general case with non-zero anomalous magnetic moments. 
Small paramagnetic pion--vector-meson loop effects are discussed in~Sec.~\ref{s:loop}.
In Sec.~\ref{s:quark}, we incorporate  the quark degrees of freedom. The results obtained within the
quark-meson framework, including the necessary vacuum effects, are favorably compared to the lattice data for $\chi_{B}$ at temperatures from $T=135$~MeV 
(the limit from the available lattice data for $\chi_{\cal{BB}}$ and $\chi_{\cal{BS}}$) up to $T_c$. 
More technical issues, in particular, the derivation of the vacuum quark-loop or the pion--vector-meson loop contributions to $\chi_B$,
are given in the Appendices.

\section{Lattice QCD results for susceptibilities \label{s:susc_res}}

We begin by discussing the lattice data for the charge (baryon number and strangeness) and magnetic susceptibilities 
used in this work, and their comparison with the HRG model calculations. We defer the precise
definitions and derivations to the following sections.

The lattice data for $\chi_{\cal{BB}}(T)$ and $\chi_{\cal{BS}}(T)$ taken from~\cite{Bollweg:2021vqf} are
shown in Fig.~\ref{fig:fittedQuarkMass} of Sec.~\ref{s:mqT}, and are used later in this paper (the 
error bars of the data are smaller than the size of the dots). The corresponding HRG results are provided in~\cite{Bollweg:2021vqf}, as well as in other lattice references~\cite{Bazavov:2011nk,HotQCD:2012fhj,Bollweg:2021vqf}, pointing out a seamless agreement for temperatures all the way 
up to $T_c \approx 155$~MeV. Above $T_c$, as expected, the HRG results depart from the lattice data, as the hadronic degrees of freedom are no longer credible there.
 
\begin{figure}
    \centering
    \includegraphics[width=0.47\textwidth]{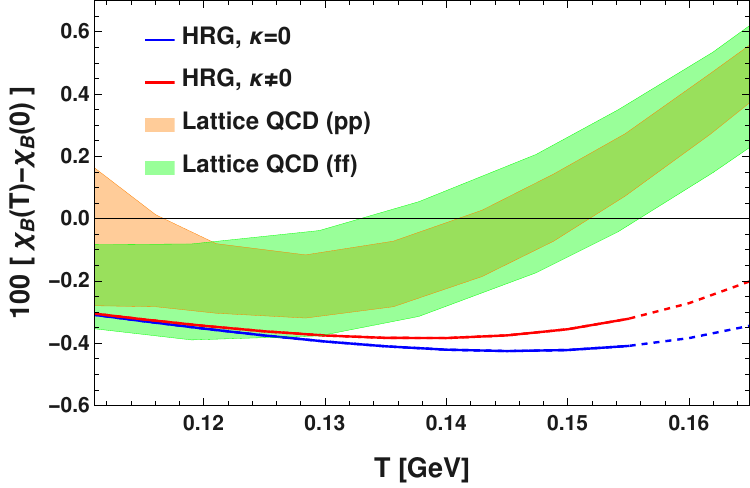}
    \caption{Magnetic susceptibility as a function of temperature, calculated in the HRG model including (red) and excluding (blue) the anomalous magnetic moments of hadrons. The lines above $T_c$ are dashed. The orange and green bands represent the lattice data obtained using two recent methods: the photon polarization method (pp)~\cite{Bali:2020bcn} and the full field method (ff)~\cite{Brandt:2024blb}, respectively.~\label{fig:chiBkap}}
\end{figure}

The results for $\chi_B(T)$ are displayed in Fig.~\ref{fig:chiBkap}. We show the two most recent 
determinations from~\cite{Bali:2020bcn} (the photon polarization method) and~\cite{Brandt:2024blb} (the full field method), indicated with the green and orange bands, respectively.
The HRG curve touches the lower-temperature (diamagnetic) end of the lattice data, where it is largely dominated by the pions, the lightest hadrons. 
This feature was originally brought to notice in~\cite{Bali:2014kia}, where for the first time the lattice simulations yielded a negative (diamagnetic) susceptibility 
at low $T$  and exhibited compatibility with HRG for $T\lesssim 120$~MeV. This fact was confirmed later in~\cite{Bali:2020bcn,Bollweg:2021vqf}. 
As discussed in the Introduction, the methodology applied in~\cite{Bali:2020bcn,Brandt:2024blb} has improved on the earlier 
results (cf.~\cite{Bonati:2013vba} and 
the comparison in Fig.~9 in~\cite{Bali:2014kia}), where $\chi_B$ was actually positive in the probed range of $T>100$~MeV.  
Importantly, as noted in~\cite{Bali:2014kia}, for $T\gtrsim 120$ MeV, HRG falls noticeably below the lattice, 
remaining large and negative near $T_c$, where 
the lattice $\chi_B$ crosses zero, changing the behavior from diamagnetic to paramagnetic.

Our HRG calculation is displayed in Fig.~\ref{fig:chiBkap}. In our study, we use QMHRG2020 -- an augmented list of hadrons including 
some additional quark-model states, 
that properly reproduces the conserved charge susceptibilities at $B=0$~\cite{Bollweg:2021vqf,Ding:2023bft,Samanta:2025mrq}. 
For a detailed discussion of variants of HRG see, e.g.,~\cite{Pal:2023sei}. We also use physical magnetic moments for hadrons, as 
explained in~\cite{Samanta:2025mrq}, which induces some more paramagnetism (cf.~the red and blue curves in Fig.~\ref{fig:chiBkap}) and raises 
the model curve by 5-10\%, still falling short of the lattice data. 

One should note that the lattice data shown in Fig.~\ref{fig:chiBkap} involve a rather large extrapolation to the physical point of $m_\pi=140$~MeV. While the mismatch between the HRG model and the lattice data displayed in~Fig.~\ref{fig:chiBkap} remains within the $2\sigma$ uncertainty, it is persistent in the wide range of $T\gtrsim 120$~MeV, indicating a genuine effect. 

We also stress that $\chi_B$, being  the curvature of the pressure (the partition function) at $B=0$ (see Sec.~\ref{s:theory}), is a low-$B$ feature. Therefore, its evaluation in the hadronic phase avoids the complications due to a possible modification of the hadronic structure and the resonances
at strong $B$~\cite{Marczenko:2024kko,Vovchenko:2024wbg}, casting doubts on the validity of HRG~\cite{Huovinen:2025nwl} in that case.
The failure of reproducing such a basic quantity as $\chi_B$ within HRG, at temperatures where it has been traditionally used in numerous applications,  
urges one to rethink the composition of the hadronic medium below $T_c$, in search of a paramagnetic component.
Moreover, the vacuum contributions, generically explicit in $\chi_B$~\cite{Endrodi:2013cs} but not treated in HRG, should 
be considered (see Sec.~\ref{s:selfcons}).

\section{Thermodynamics in a magnetic field\label{s:theory}}

\subsection{Basic definitions \label{s:def }}

Thermodynamical properties of a system treated via the grand canonical ensemble can be obtained from the 
the grand thermodynamical potential density $\Omega$~\cite{Landau:1980mil,pathriaStatisticalMechanics2011}, which in the presence of a magnetic field is
\begin{eqnarray}
\Omega=U - TS -B {\cal M}-\mu N, \label{eq:omega}
\end{eqnarray}
where $U$ is the internal energy, $S$ - the entropy, ${\cal M}$ - magnetization, and $\mu N=\mu_{\cal B} {\cal B}+\mu_S {\cal S}+...$ involves the combination of chemical potentials and the corresponding conserved charges.\footnote{To distinguish the baryon number from the magnetic field, we use 
${\cal B}$ for  the former, and $B$ for the latter.} 
In the thermodynamic limit of a large volume, $\Omega$ is
related to the system pressure, $P=-\Omega/V$.

The conserved charge susceptibilities used in this work are defined as 
\begin{eqnarray}
\chi_{\cal BB}=\left . \frac{1}{T^2} \frac{\partial^2 P}{\partial \mu_{\cal B}^2} \right |_{\mu_{\cal B,S}=0}, \;\;\; 
\chi_{{\cal B}S}=\left . \frac{1}{T^2} \frac{\partial^2 P}{\partial \mu_{\cal B}\partial \mu_{\cal S}} \right |_{\mu_{\cal B,S}=0}.
\label{eq:defchiBBchiBS}
\end{eqnarray}
They are evaluated at vanishing chemical potentials, as in lattice QCD.

According to Eq.~(\ref{eq:omega}), the magnetization density of the system is obtained as
 \begin{eqnarray}
    {\mathcal M}=\frac{\partial \mathcal{P}}{\partial B},
    \label{eq:defmagnetization}
\end{eqnarray}
while the magnetic susceptibility is given by the second derivative of the pressure at $B=0$,
 \begin{eqnarray}
     \chi_B=\frac{\partial \mathcal M}{\partial B} \bigg|_{B=0}=\frac{\partial^2 P}{\partial B^2}\bigg|_{B=0}.
     \label{eq:defchiB}
\end{eqnarray}

Generically, the pressure can be decomposed into the thermal and vacuum parts, $P=P^{\rm th}+P^{\rm vac}$.
In the non-interacting particle picture, such as HRG or the quark-meson model studied later, at $B=0$ the thermal part for
an individual particle species (for brevity, we omit indexing of the particle species below) has the form 
\begin{eqnarray}
    P^{\rm th}= -\eta T \sum_{s_z} \int \frac{d^3p}{(2\pi)^3} \log[1-\eta f(E_p,T,\mu)], \label{eq:ther}
\end{eqnarray}
where $f(E_p,T,\mu)$ represents the occupation number of a stable hadron or a resonance with a given statistics,
\begin{eqnarray}
f(E_p,T,\mu) = \frac{1}{\exp(\frac{E_p-\mu}{T})+ \eta}.
\label{eq:distfunc}
\end{eqnarray}
Here $E_p=\sqrt{m^2+p^2}$ denotes the energy, $s_z$ is the spin projection, and $\eta=+1$ or $-1$ depending on whether the particle is a fermion or a boson. 
The corresponding vacuum part can be formally written as~\cite{Endrodi:2013cs}
\begin{eqnarray}
 P^{\rm vac}=-\tfrac{1}{2} \eta  \sum_{s_z} \int \frac{d^3p}{(2\pi)^3} E_p, \label{eq:vacu}
\end{eqnarray} 
which is infinite, thus requires regularization, to be discussed later and detailed in Appendix~\ref{s:vacu_quark}.

In the presence of a magnetic field $\vec{B} = B \hat{z}$, for charged particles the corresponding formulas involve a sum 
over the Landau levels $l$~\cite{Landau:1980mil,pathriaStatisticalMechanics2011}
 \begin{eqnarray}
&& P^{\rm th}= -\eta T \frac{B|Q|}{2\pi^2} \sum_{l=0}^\infty \sum_{s_z} \int_0^\infty dp_z \log[1-\eta f(p_z,B,l)], \nonumber \\
&& P^{\rm vac}=  -\eta \frac{B|Q|}{4\pi^2} \sum_{l=0}^\infty \sum_{s_z} \int_0^\infty dp_z E(p_z,B,l).
\label{eq:pressurefull}
\end{eqnarray}
For the neutral particles, equations of the form (\ref{eq:ther},\ref{eq:vacu}) hold, but $E$ may 
explicitly depend on $B$ if the particle has an anomalous magnetic moment (cf. Sec.~\ref{s:HRG_basics}).

\section{Hadron Resonance Gas  \label{s:HRG}}

\subsection{Energy spectra \label{s:HRG_basics}}

In HRG, one deals with a non-interacting system of stable hadrons and resonances, which incorporate in a simplified way the interactions between the stable 
hadrons~\cite{Fermi:1950jd,Pomeranchuk:1951ey,Hagedorn:1965st,Koch:1986ud,Dashen:1969ep,Venugopalan:1992hy,Broniowski:2001we,Huovinen:2009yb}. 

The energy spectra of individual hadrons in a magnetic field depend on their spin. The relevant formulas were reviewed in~\cite{Samanta:2025mrq}.
In particular, for spin-0 charged states (e.g., $\pi^\pm$, $K^\pm$),
\begin{eqnarray}
\hspace{-4mm} E(p_z,B,l)\equiv\epsilon_{l,p_z}=\sqrt{m^2+p_z^2 + B |Q| (2l+1)},\;
\label{eq:enSpin0}
\end{eqnarray}
where $Q$ denotes the particle electric charge and $l$ labels the Landau level.

For spin-$\tfrac{1}{2}$ particles with an  anomalous magnetic moment  $\kappa$, one has the Tsai-Yildiz formula~\cite{Tsai:1971zma}
\begin{eqnarray}
&&\epsilon_{l,p_z}=\nonumber \\
&&\sqrt{\left (\sqrt{m^2+ B \left[|Q|(2l\!+\!1\!)-\!2Q s_z\right]}-2\mu_M B\kappa s_z\right)^2\!+\!p_z^2}\nonumber \\
\label{eq:TsaiYldiz}
\end{eqnarray}
for states of non-zero charge $Q$ (such as the proton), and 
\begin{eqnarray}
&&\hspace{-10mm} \epsilon_{p,p_z}=\sqrt{\left (\sqrt{m^2+ p^2-p_z^2}-2\mu_M B\kappa s_z\right)^2+p_z^2}  
\end{eqnarray}
for neutral states (such as the neutron).  Here $\mu_M=\frac{1}{2m}$ is the natural magneton of the particle. Our 
convention for the gyromagnetic ratio of the hadron is
\begin{eqnarray}
g=2(Q+\kappa). \label{eq:gyro}
\end{eqnarray}

For higher spin states we adopt the same prescription as in~\cite{Samanta:2025mrq}, where the
anomalous magnetic moments of hadrons are treated  non-relativistically as small corrections. For instance, 
in case of spin-1 or spin-$\tfrac{3}{2}$ (Rarita-Schwinger) charged states~\cite{dePaoli:2012eq} (e.g., $\rho$ or $\Delta$), we use~\cite{Samanta:2025mrq}
\begin{eqnarray}
&&\epsilon_{l,p_z}=\sqrt{m^2+p_z^2+B [|Q|(2l\!+\!1\!)-\!2Q s_z]}- 2 B \mu_M \kappa s_z,  \nonumber \\
\label{eq:enDelta}
\end{eqnarray}
and for particle with spin larger than $\tfrac{3}{2}$, we treat the entire spin interaction non-relativistically~\cite{Samanta:2025mrq},
\begin{eqnarray}
&&\hspace{-4mm} \epsilon_{l,p_z}=\sqrt{m^2+p_z^2+B |Q|(2l\!+\!1\!)}-  g B \mu_M  s_z .
\label{eq:enHighspin}
\end{eqnarray}

With the above spectra one can obtain the pressure for individual hadron species~(\ref{eq:pressurefull}), which then  
can be used to obtain the susceptibilities. 
The values of $\kappa$ used in the calculations presented 
in this paper are taken from Table~I compiled in~\cite{Samanta:2025mrq}.

\subsection{Baryon-baryon and baryon-strangeness susceptibilities \label{s:HRG }}

Using Eq.~(\ref{eq:defchiBBchiBS}), one obtains a generic expression for $\chi_{Q_1Q_2}$ from 
the thermal part of the pressure\footnote{It simply follows from the 
fact that $f (1 - \eta f)= -T df/d{E_p}$ and an integration by parts.}  
\begin{eqnarray}
&&\hspace{-6mm}\chi_{Q_1 Q_2}=  \frac{Q_1 Q_2(2s+1)}{2\pi^2 T^2} \int_0^\infty \!\! dp\;  \left(m^2+2p^2\right) \frac{f(E_p)}{E_p},
\end{eqnarray}
where $Q_i={\cal B}$ or $S$, $s$ is the particle's spin, and $f(E_p)\equiv f(E_p,T, \mu=0)$.
Summing over the particle species yields $\chi_{\cal BB}$ and $\chi_{{\cal B}S}$ in HRG, shown in Figs.~2 and 3 of~\cite{Bollweg:2021vqf}. The corresponding lattice data are shown in Fig.~\ref{fig:fittedQuarkMass}. 
In~\cite{Bollweg:2021vqf}, it was noted that HRG is successful in describing the corresponding lattice data both 
qualitatively and quantitatively up to $T_c\approx 155$~MeV. Above $T_c$, one observes a gradual deviation, typically attributed to an emergence of the QCD degrees of freedom.

\subsection{Magnetic susceptibility with anomalous magnetic moments \label{s:HRG_magsusc}}

In HRG, one considers only the thermal part of the magnetic susceptibility. 
Some details of the straightforward derivations of the following formulas are given in Appendix~\ref{s:chiB_derv}). 

For the diamagnetic component one gets,
\begin{eqnarray}
 \chi_B^{\rm dia} = -\frac{Q^2(2s+1)}{24\pi^2} \int_0^\infty dp \; \frac{f(E_p)}{E_p},
 \label{eq:chiBdia}
\end{eqnarray}
which is clearly negative-definite and monotonic for each particle.

The form of the paramagnetic component depends on the spectrum used. 
For spin-$\tfrac{1}{2}$ particles with the Tsai-Yildiz expression~(\ref{eq:TsaiYldiz}), one gets (see Appendix~\ref{s:chiB_derv})
\begin{eqnarray}
    &&\hspace{-7mm} \chi_{B;s=\frac{1}{2}}^{\rm para} = \frac{s(s+1)(2s+1)}{6\pi^2} \int_0^\infty dp \times \nonumber \\ 
    &&~~~\left[Q^2+2 Q \; \kappa + \left(1+\frac{p^2}{m^2}\right) \; \kappa^2 \right] \; \frac{f(E_p)}{E_p}.
    \label{eq:chiBparaTsaiYildiz}
\end{eqnarray}
The cases corresponding to spectra of Eqs.~(\ref{eq:enDelta},\ref{eq:enHighspin}) used for higher spin states 
are listed in Appendix~\ref{s:chiB_derv}. Obviously, all are positive-definite.
In the limit of a large mass $m$, all the expressions reduce to the formula 
\begin{eqnarray}
    &&\chi_{B;m \gg T}^{\rm para} = \frac{s(s+1)(2s+1)g^2}{24\pi^2} \int_0^\infty \!\! dp  \frac{f(E_p)}{E_p},
    \label{eq:largem}
\end{eqnarray}
becoming proportional to the gyromagnetic factor squared.
The sum $\chi_B^{\rm dia} +  \chi_B^{\rm para}$ is positive for any $s>0$ state and monotonic.

From the formulas for $\chi_B^{\rm para}$ it is clear that, as expected, the inclusion of the anomalous 
magnetic moments enhances the paramagnetic component of the magnetic susceptibility in HRG.
However, with the physical $\kappa$'s, the total enhancement effect is mild, as can be seen from the comparison in Fig.~\ref{fig:chiBkap}. 
We recall that the inclusion of the physical $\kappa$'s, which are large for baryons, 
was crucial to achieve agreement for the conserved charge susceptibilities in HRG at nonzero values 
of $B$~\cite{Samanta:2025mrq}. 
In Fig.~\ref{fig:chiBkap}, the HRG curves above $T_c$ are dashed, to indicate they only carry a formal meaning there. At lower values of $T$, the relative effect of $\kappa$ decreases, to become $\approx 15\%$ of the total $\chi_B$ at $T=140$~MeV. 
Hence, the mismatch between HRG results (obtained from the thermal part of the pressure) and the lattice persists even after including the anomalous magnetic moments. 

%We note that in Figs.~\ref{fig:chiB} and \ref{fig:chiBindv} the results labeled ``HRG'' include the anomalous magnetic moments. 

\subsection{Anatomy of the magnetic susceptibility \label{s:HRG_anatomy}}

\begin{figure}
    \centering
    \includegraphics[width=0.47\textwidth]{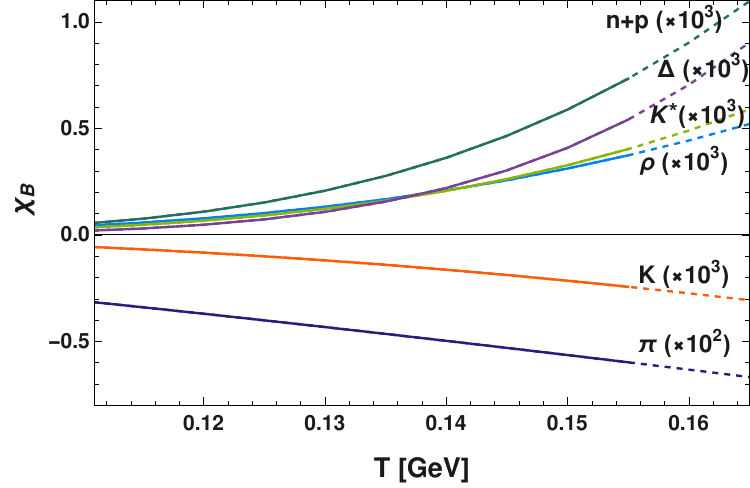}
    \caption{Magnetic susceptibility of several hadron species as functions $T$ with $\kappa \neq0$. The numbers in parentheses next to labels denote factors by which a given result is multiplied (note a different factor for the pion). The hadron labels include the antiparticles and the spin-isospin degeneracy.  \label{fig:chiBindv}}
\end{figure}

In order to better understand the HRG results, in Fig.~\ref{fig:chiBindv} we show the contributions
from the dominating hadron species plotted as functions of $T$ including their anomalous magnetic moment $\kappa$. The numbers in parentheses next to the label of each curve 
denote a factor by which the given result is multiplied, which is $10^2$ for the pion and $10^3$ for the other hadrons. 
The result for each curve includes the corresponding anti-particles, as well as the spin and isospin (charge) degeneracy.

The magnitude of $\chi_B$ for each hadron species, naturally, grows with the increasing temperature for both the paramagnetic and diamagnetic cases. 
It is controlled by the mass of the hadron through the occupation number via the formulas from Sec.~\ref{s:HRG_magsusc} and Appendix~\ref{s:chiB_derv}),
which at large masses are damped approximately as $\sim e^{-M/T}$.
Thus, at lower values of $T$ higher mass hadrons are strongly suppressed and one is left with the leading diamagnetic contribution of the pion. 
As $T$ increases, the kaons join the diamagnetic contribution. As a result, at 
temperature range $T<165~{\rm MeV}$, the nature of HRG remains diamagnetic.
At higher values of $T$ other states appear, bringing 
in paramagnetic contributions, mostly from the numerous baryonic states of higher mass~\cite{Broniowski:2004yh}, which 
actually at large $T\approx 170~{\rm MeV} >T_c$ (well outside the expected validity of HRG) prevail over the diamagnetic terms.

\subsection{Magnetic polarizabilities of hadrons \label{s:HRG_magpol}}

An external constant and uniform magnetic field induces deformation of a composite object, whose dynamical response (at $T=0$) is quantified with 
the magnetic polarizability. To quadratic order, the energy picks up a contribution
\begin{eqnarray}
    \Delta E = -\frac{1}{2} \beta B^2, \label{eq:pola}
\end{eqnarray}
where $\beta$ is the magnetic polarizability of the hadron. 
The estimates are $\beta_{\pi^\pm}\approx -2 \times 10^{-4}{\rm ~fm}^3$~\cite{Moinester:2022tba},  
$\beta_{\pi^0}\approx 3 \times 10^{-4}{\rm ~fm}^3$\cite{He:2020ysm}, 
$\beta_{n}\approx 4 \times 10^{-4}{\rm ~fm}^3$~\cite{ParticleDataGroup:2024cfk}, and $\beta_{p}\approx 2 \times 10^{-4}{\rm ~fm}^3$~\cite{ParticleDataGroup:2024cfk}.
The term (\ref{eq:pola}) in the definition of $\chi_B$ yields an extra piece for a particle species,
\begin{eqnarray}
    \chi_B^{\rm pol} = \frac{\beta}{2\pi^2} \int_0^\infty dp \;  p^2 \; f(E_p),
\end{eqnarray}
which, as expected, is proportional to the density of states at $B=0$. With the above-quoted values for $\beta$'s, we find that $100 \chi_B^{\rm pol} \approx -0.002$, 
negligible compared to the values of Fig.~\ref{fig:chiBindv}.

\subsection{Magnetization \label{s:HRG_magnetization}}

\begin{figure}
    \centering
    \includegraphics[width=0.47\textwidth]{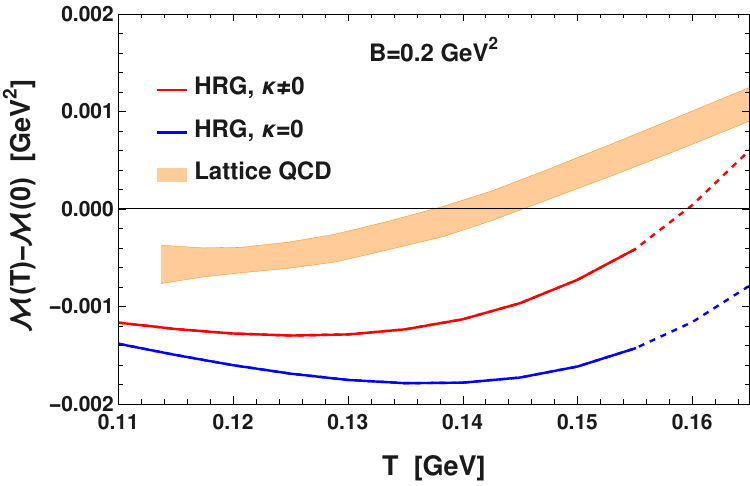}
    \caption{Subtracted magnetization,
    $\mathcal{M}(B,T)-\mathcal{M}(B,0)$, at $B=0.2~{\rm GeV}^2$, plotted as a function of $T$. 
    The lattice result, obtained from~\cite{Bali:2014kia}, is represented with the orange band. The blue and red curves represent, respectively, 
    the HRG results without and with the anomalous magnetic moments of hadrons.  \label{fig:chiM}}
\end{figure}

To corroborate the qualitative findings for $\chi_B$, we also examine the magnetization $\mathcal{M}$ of Eq.~\ref{eq:defmagnetization}. 
This quantity depends on the magnitude of the magnetic field $B$, which here we set to $B=0.2~{\rm GeV}^2$, sizable but not too large to avoid the 
problems for hadrons in a too strong magnetic field~\cite{Vovchenko:2024wbg,Marczenko:2024kko,Samanta:2025mrq}. Since $\mathcal{M}$ needs 
subtraction due to the vacuum contribution~\cite{Endrodi:2013cs},
we plot this quantity subtracted at $T=0$, namely $\mathcal{M}(B,T)-\mathcal{M}(B,0)$. Note that the possible vacuum contribution 
is absent in HRG, similar to the case of $\chi_B$.

Figure~\ref{fig:chiM} shows the results for the subtracted magnetization in HRG compared with lattice data~\cite{Bali:2014kia}. 
We observe a qualitatively similar feature as in $\chi_B$, namely a too strong diamagnetism in HRG, placing the model curve way below the data. 
The effect of the anomalous magnetic moments of hadrons on $\mathcal{M}(B,T)-\mathcal{M}(B,0)$ is visible.

\section{Pion--vector-meson loops \label{s:loop}}

We now turn to yet another contribution to the paramagnetism of the 
medium, which originates from the pion--vector-meson loops. While such a contribution is not 
immediately apparent in HRG, it is manifested when considering  the photon polarization in a hadronic medium, from which $\chi_B$ can be obtained.
The possibility of couplings between photons, pions, and vector mesons involving the Levi-Civita tensor $\epsilon^{\alpha \beta \mu \nu}$ is well known,
and such ``anomalous'' effects are expected to be generically small. Yet, it is worthwhile to examine the issue in a great detail to have a complete 
hadronic picture.

\begin{figure}
\centering
\includegraphics[width=0.3\textwidth]{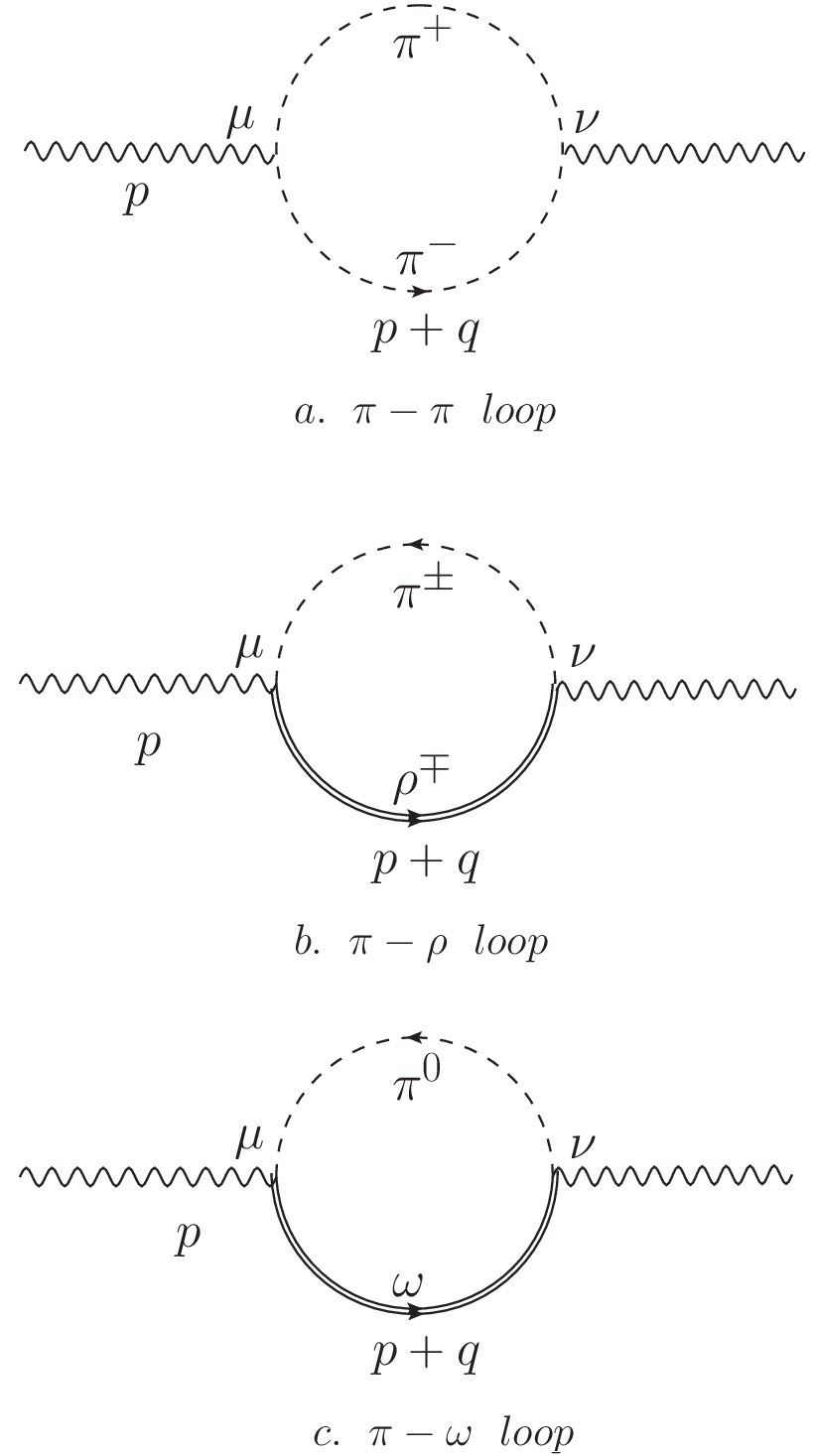}
\caption{Vacuum polarization diagrams for photons, involving a $\pi$ loop (a) and  $\pi$--vector-meson loops, (b) and (c).}
\label{fig:pi-V loop}
\end{figure}

According to general considerations~\cite{Weldon:1982aq,Fujimoto:1988ms,Endrodi:2013cs,Endrodi:2022wym},
\begin{eqnarray}
    \chi_B = \frac{1}{2} \lim_{q\rightarrow 0} \lim_{q_0\rightarrow 0} \frac{\partial^2 \text{Re}[\Pi_s] }{\partial q^2}.
    \label{eq:defchiB_vacpol_thermal}
\end{eqnarray}
To pick up the thermal contribution, one
calculates the vacuum polarization tensor of the photon using the loop diagrams of~Fig.~\ref{fig:pi-V loop} where one of the propagators in the loop
corresponds to the thermal part. For the case where both propagators in the loop correspond to the same particle, such as for instance the pion,
one obtains the formula of Eq.~(\ref{eq:chiBdia}) as a check.

The loops, however, can mix different states. In particular, there is the above-mentioned coupling between a 
photon, pion, and a vector meson (cf.~Fig.~\ref{fig:pi-V loop}). A detailed calculation of these
diagrams is provided in Appendix~\ref{s:pi-rho}. The magnetic susceptibility in this case becomes,
\begin{eqnarray}
    \chi_B = -\frac{1}{12\pi^2\Delta}\left(\frac{g_{\gamma \pi V}^2}{m_V^2} \right) \left[ I_\pi -I_V\right], \label{eq:ipi}
    \label{eq:chiBpirho}
\end{eqnarray}
where $V=\rho, \omega$, $\Delta = m_V^2-m_\pi^2$, $g_{\gamma \pi V}$ is the coupling constant, and $I_{a}$ is given by,
\begin{eqnarray}
    I_{a} = \int p^2 dp \frac{f(E_{a})}{E_{a}} \left(p^2-3E_{a}\right), \;\;\; a=\pi,V.
\end{eqnarray}
Here $E_{a}=\sqrt{p^2+m_a^2}$ represents the energy of meson $a$. 
The values of the coupling constants have been determined from the vector meson electroproduction data and have the values~\cite{Oh:2000pr,Gokalp:2001sr}
\begin{eqnarray}
g_{\gamma \pi \rho} = 0.54, \;\;\; g_{\gamma \pi \omega}= 1.82.
\end{eqnarray}
In Eq.~(\ref{eq:ipi}), $I_\pi$ dominates over $I_V$ due to the small mass of the pion compared to the masses of $\rho$ or $\omega$.

\begin{figure}
    \centering
    \includegraphics[width=0.47\textwidth]{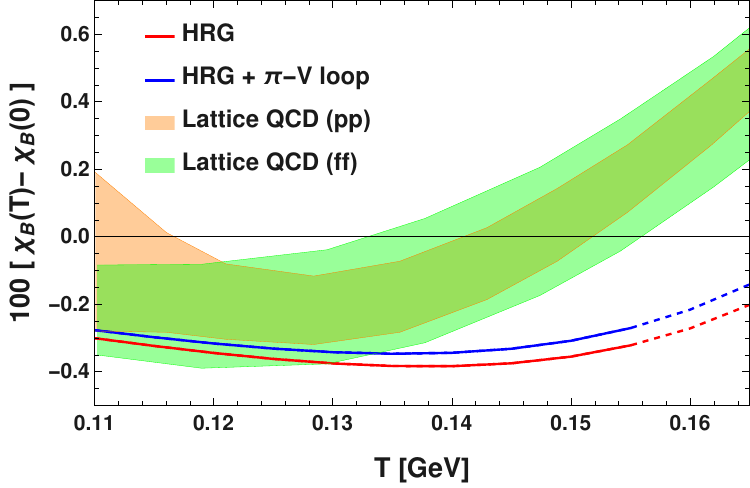}
    \caption{Magnetic susceptibility as a function of temperature in usual HRG (red) and HRG with the contribution coming 
    from anomalous coupling such as $\rho-\pi$ and $\omega-\pi$ loops in the vacuum polarization of $\gamma$ (blue), compared with lattice data (orange and green).}
    \label{fig:chiBhadronmix}
\end{figure}

Fig.~\ref{fig:chiBhadronmix} shows that the contribution of the pion--vector-meson loops to $\chi_B$ is indeed small. 
To assess the effect, we use the HRG result from Fig.~\ref{fig:chiBkap}. It can be seen that the mixing effect is strictly paramagnetic and moderate, 
ranging from $10\%$ to $20\%$ while going from low to 
high temperatures. Nevertheless, it should be 
noted that the $\pi-\rho$ contribution is larger than the $\rho-\rho$ contribution, hence carries some significance. 
Notably, the sign and the order of magnitude of the enhancement from the 
pion--vector-meson loops is in a quantitative agreement with a two-loop chiral perturbation theory calculation~\cite{Hofmann:2021bac}.

\section{Quark degrees of freedom \label{s:quark}}

\subsection{Motivation}

In the previous sections we have verified that within HRG it is not possible to achieve agreement with the lattice results 
for $\chi_B(T)$ in the fiducial temperature range. Even after including the anomalous magnetic moments of the hadrons
or the pion--vector-meson loops, the model remains unsuccessful in describing the lattice data.
A general conclusion is that one needs sufficiently low paramagnetic states (with $s>0$ and of sufficiently low mass), such that at $T \approx T_c$ their paramagnetism may overcome the diamagnetism from the pions and kaons. 
Obvious candidates here are the (constituent) quarks, which as spin-$\tfrac{1}{2}$ states are paramagnetic and light enough. 
Moreover, the inclusion of quarks allows one to model the needed vacuum term~\cite{Endrodi:2013cs}, advocated in Sec.~\ref{s:selfcons}.

As seen from the lattice data, we also need 
to retain the diamagnetism at low values of $T$, where $\chi_B(T)$ is negative. For that reason we shall keep the pions and kaons
as additional degrees of freedom, in the spirit of the quark-meson models (see~\cite{Schaefer:2004en} and references therein). 
Thus, we consider an independent particle model with constituent quarks of three colors and three flavors, pions, and kaons. 
Recently, quark meson models at non-zero $B$ have been in similar tasks in~\cite{Kamikado:2014bua,Mao:2024gox,Mei:2024rjg}.

\subsection{Temperature-dependent quark masses from fits to $\chi_{BB}$ and $\chi_{BS}$ \label{s:mqT}}

\begin{figure}
    \centering
    \includegraphics[width=0.45\textwidth]{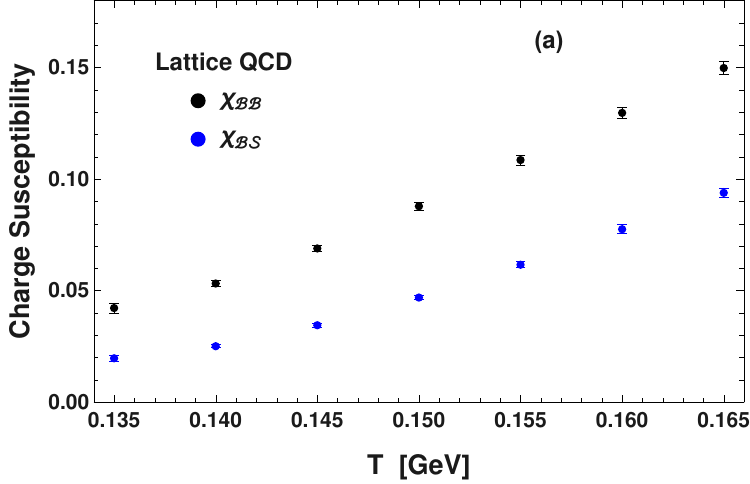}\\
    ~~\includegraphics[width=0.47\textwidth]{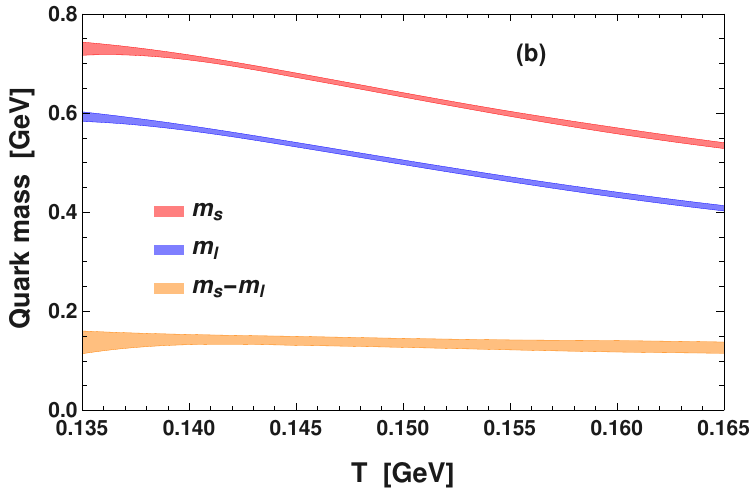}
    \caption{(a) Lattice data~\cite{Bollweg:2021vqf} for susceptibilities $\chi_{\cal{BB}}$ (black) and $\chi_{\cal{BS}}$ (blue) as functions of $T$. (b) Light and strange quark masses as functions of $T$ from the fit to the lattice data for $\chi_{\cal BB}$ and $\chi_{{\cal B}S}$ in panel (a) 
    (the blue and red bands, respectively), and the difference $m_s-m_l$ (orange band). The width of the bands reflects the uncertainty of the lattice data.\label{fig:fittedQuarkMass}}
\end{figure}

We fix the quark masses as functions of $T$ from the requirement to describe the lattice data for $\chi_{\cal BB}$ and $\chi_{{\cal B}S}$ from~\cite{Bollweg:2021vqf}
shown in  panel~(a) of Fig.~\ref{fig:fittedQuarkMass}. We use here Eqs.~(\ref{eq:selfsus}) and (\ref{eq:selfsusBS}) and 
consider equal mass for $u$ and $d$ quarks, denoted by $m_l=m_u=m_d$, and an independent strange quark mass $m_s$. 
The procedure goes as follows: First, $m_s(T)$ is fitted to the $\chi_{{\cal B}S}$ data at each value of $T$, according to Eq.~(\ref{eq:selfsusBS}) where only 
the strange quarks contribute. 
Then, we use thus obtained $m_s(T)$ to fix $m_l(T)$ with the $\chi_{\cal BB}$ data, according to Eq.~(\ref{eq:selfsus}), where the quarks of all three flavors contribute.

The above methodology is analogous to the one from~\cite{Mykhaylova:2019wci}, where the entropy density was used to find the $T$-dependence
of constituent quarks and gluons, termed as quasiparticles~\cite{Chakraborty:2010fr}.

The obtained 
result for the quark masses is presented in Fig.~\ref{fig:fittedQuarkMass}. First, we note that at $T\approx T_c$ they 
are significantly larger than the traditional constituent masses, where $m_l \approx 300$~MeV, but, interestingly, 
are in the ballpark of the quasiparticle masses of Fig.~2 in~\cite{Mykhaylova:2019wci}, where
$m_l(T_c)\approx 550$~MeV and $m_s(T_c)\approx 600$~MeV. Second, the splitting $m_s-m_l$ in our Fig.~\ref{fig:fittedQuarkMass} 
is remarkably stable in the displayed 
range of $T$, being around 120--130~MeV, which is a typically expected value in the constituent quark picture.

\subsection{Vacuum contribution \label{s:vac}}

\begin{figure}[b]
\centering
\includegraphics[width=0.32\textwidth]{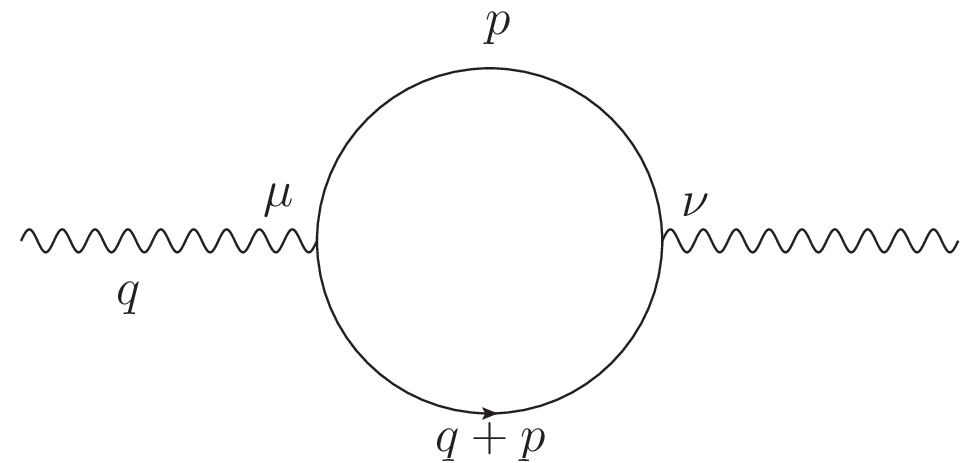}
\caption{Photon vacuum polarization diagram for quarks}
\label{fig:quark_loop}
\end{figure}

As discussed in Sec.~\ref{s:selfcons}, the vacuum contribution to the pressure depends explicitly on $B$ via the energy spectrum, 
therefore making it important for the evaluation of $\chi_B$. 
In the quark-meson model, the vacuum terms were computed in~\cite{Kamikado:2014bua}, with dimensional regularization and renormalization applied. 
For the quarks, the vacuum contribution to $\chi_B$ can be evaluated according to the diagram of Fig.~\ref{fig:quark_loop}, 
representing the photon polarization in the quark model, $\Pi^{\mu \nu}$. 
We use the formula (see, e.g.,~\cite{Weldon:1982aq,Fujimoto:1988ms,Endrodi:2013cs,Endrodi:2022wym}) 
\begin{eqnarray}
    \chi_B^{q, {\rm vac}} = \frac{1}{2} \lim_{\vec{q}\rightarrow 0} \lim_{q_0\rightarrow 0} \frac{\partial^2 \text{Re}[\Pi_s^{q, {\rm vac}}] }{\partial q^2},
    \label{eq:defchiB_vacpol}
\end{eqnarray}
with $q_0$ and $\vec{q}$ denoting the photon's momentum and $\Pi_s^{q, {\rm vac}}=\tfrac{1}{2} \sum_{i=1}^3\Pi_{ii}^{q, {\rm vac}}$.

Here, in the spirit of effective quark models, rather than carrying out the renormalization procedure, we 
apply a regulator and keep it to remove the hard-momentum contributions. Specifically, 
we use the Pauli-Villars regularization scheme with two-subtractions~\cite{RuizArriola:2002wr,Broniowski:2021awb}, as given by Eq.~(\ref{eq:PV}).
We also incorporate the possible anomalous magnetic moments of the 
constituent quarks, taking the vertex in the form~\cite{Tsai:1971zma,Bicudo:1998qb,Xu:2020yag,Ghosh:2021dlo}  
\begin{eqnarray}
\Gamma^\mu=Q\gamma^\mu+\frac{\kappa}{2m} \sigma^{\mu\nu}q^\nu
\end{eqnarray}
(see Appendix~\ref{s:vacu_quark} for details).

With the described procedure we find explicitly for each quark species (treating color separately)
\begin{eqnarray}
     &&\chi_B^{q, {\rm vac}}(T) = \frac{Q^2}{36 \pi^2} \left( \frac{2\Lambda^4+3\Lambda^2m^2}{(\Lambda^2+m^2)^2}-3\log[\frac{\Lambda^2}{m^2}+1]\right) \nonumber \\
     &&\hspace{13mm}+\frac{\kappa\; Q}{4\pi^2} \left( \log[\frac{\Lambda^2}{m^2}+1] - \frac{\Lambda^2}{\Lambda^2+m^2}\right) \nonumber \\
     &&\hspace{1mm}+ \frac{\kappa^2}{48 \pi^2} \bigg( \Lambda^2 (\frac{4}{\Lambda^2+m^2}-\frac{1}{m^2})- 3\log[\frac{\Lambda^2}{m^2}+1]  \bigg),
     \label{eq:chiB_vac}
\end{eqnarray}
where $\Lambda \approx 800$~MeV (such values are typically obtained when considering the pion 
structure in the vacuum~\cite{RuizArriola:2002wr}) is the scale of the regulator, while $m=m(T)$ is the 
temperature-dependent quark mass.\footnote{For the $\kappa=0$ case, 
taking the limit $\Lambda\rightarrow\infty$ and subtracting the $T=0$ term yields the same renormalized formula as in~\cite{Kamikado:2014bua}.}

\begin{figure}
    \centering
    \includegraphics[width=0.47\textwidth]{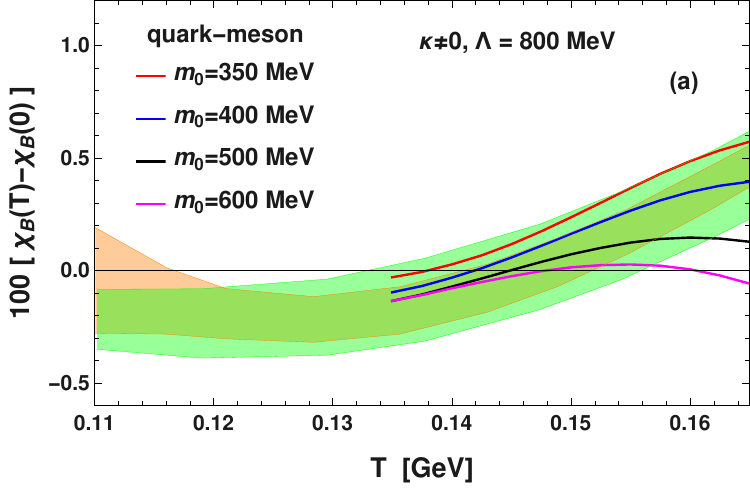}\\
    \includegraphics[width=0.47\textwidth]{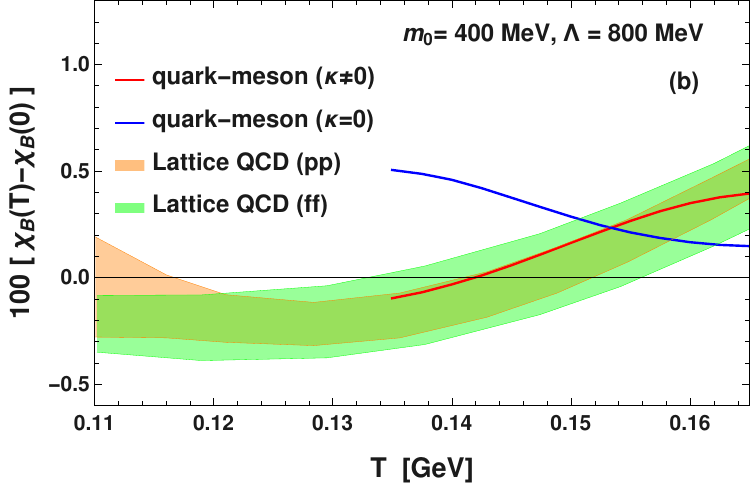}\\
    \includegraphics[width=0.47\textwidth]{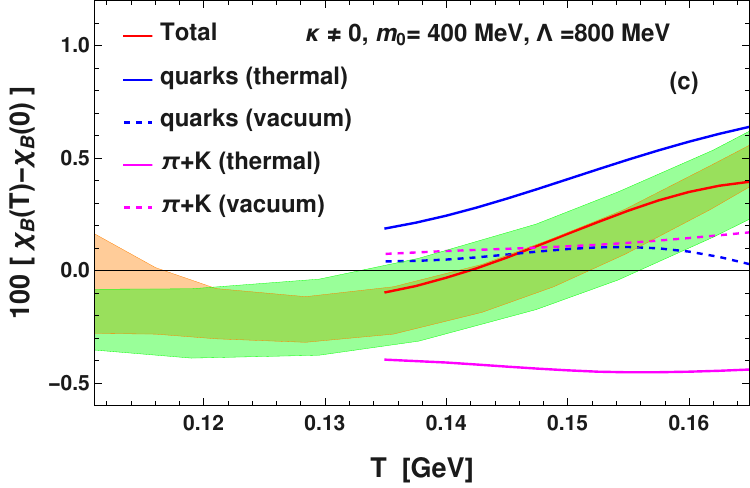}\\
    \caption{(a)~Magnetic susceptibility as a function of $T$ in the quark-meson model 
    with the fitted quark masses from Fig.~\ref{fig:fittedQuarkMass} and the temperature dependent meson masses 
    taken from~\cite{Kamikado:2014bua}. The results include both the 
    thermal and vacuum parts with non-zero quark anomalous magnetic moments. The results are 
    plotted for several values of light quark mass $m_{l}(0)\equiv m_0$
    and $m_s(0)=m_l(0)+120$~MeV. (b)~The effect of anomalous magnetic moments of the quarks in the quark-meson model. 
    (c)~Anatomy of $\chi_B(T)-\chi_B(0)$  for $m_l(0)=400$~MeV, $\kappa_q \neq 0$ and $\Lambda=800$~MeV \vspace{1cm}\label{fig:chiBwithvacuum}.}
\end{figure}

For the mesons, one can  compute the pion (or kaon) loop in the photon vacuum polarization diagram (see Fig.~\ref{fig:pi-V loop}). 
The leading term, renormalized to 0 at $T=0$, is~\cite{Kamikado:2014bua} 
\begin{eqnarray}
 \hspace{-5mm}   \chi_{B}^{\rm mes, vac}(T)- \chi_{B}^{\rm mes, vac}(0) = - \frac{1}{48 \pi^2} \log\left[\frac{m(0)}{m(T)}\right],
    \label{eq:chiB_mesons}
\end{eqnarray}
where $m(T)$ denotes the temperature dependent pion or kaon mass. Numerous lattice and model studies suggest 
that these masses grow very slowly up to 
$\approx T_c$, where 
they start to increase rapidly.

\subsection{Quark anomalous magnetic moments  \label{s:qan}}

The quark anomalous magnetic moments are estimated from a non-relativistic quark model calculation of the baryon magnetic moment $\mu_p$, $\mu_n$, 
and $\mu_{\Lambda^0}$, using~\cite{Thirring:1965,Karl:1991hi,Capstick:1996ib,Bicudo:1998qb}
\begin{eqnarray}
    &&\mu_p = \frac{4}{3} \mu_u - \frac{1}{3} \mu_d, \nonumber\\
    &&\mu_n = \frac{4}{3} \mu_d - \frac{1}{3} \mu_u, \nonumber\\
    &&\mu_{\Lambda} =\mu_s.
     \label{eq:kappa_quark}
\end{eqnarray}
We solve the above equations with the physical values of the baryon magnetic moments (collected,
e.g., in Table~I of~\cite{Samanta:2025mrq}). The result is,  $\mu_u=1.85 \; \mu_N$,   $\mu_d= -0.97 \; \mu_N$, and  $\mu_s=-0.61 \; \mu_N$, which are assumed to be independent of $T$. We then apply the basic equality, $\mu=g s/2m$, where $s$ is the spin and $m$ the mass, and using Eq.~(\ref{eq:gyro}) we get the anomalous magnetic moments $\kappa_q$.

\subsection{Magnetic susceptibility \label{s:quark_magsusc}}

The full magnetic susceptibility in the quark-meson model is composed as follows:
\begin{eqnarray}
&& \chi_B(T)=N_c \sum_{u,d,s} \left [ \chi_{B}^{\rm q, th}(T) + \chi_{B}^{\rm q, vac}(T) \right ]  \nonumber \\ 
&& + \sum_{\pi, K} \left [ \chi_{B}^{\rm mes, th}(T) + \chi_{B}^{\rm mes, vac}(T) \right ]. \label{eq:chiB_quarks}
\end{eqnarray}
In this formula, the quark and meson masses are temperature-dependent. For the quark case, we take the dependence found in Sec.~\ref{s:mqT} in the available 
range of $T$, while for $m_\pi(T)$ and $m_K(T)$ we use the model results from the quark-meson model given in Fig. 3 of~\cite{Kamikado:2014bua}. 

Since the lattice data~\cite{Endrodi:2022wym,Brandt:2024blb} correspond to the subtraction $\chi_B(T)-\chi_B(0)$, we do the same in the quark meson model.
The thermal parts vanish at $T=0$, the vacuum meson part of Eq.~(\ref{eq:chiB_mesons}) is already subtracted, whereas for the 
subtraction of the quark vacuum part we would need the quark masses at $T=0$. However, these are inaccessible to the method of Sec.~\ref{s:mqT}, since the lattice data 
for $\chi_{\cal BB}$ and $\chi_{{\cal B}S}$ at $T=0$ are not available. 
Thus, we first probe several reasonable values for $m_l(0)$ and $m_{s}(0)=m_{l}(0)+120$~MeV, and later use a different subtraction scheme.
  
\begin{figure}
    \centering
    \includegraphics[width=0.47\textwidth]{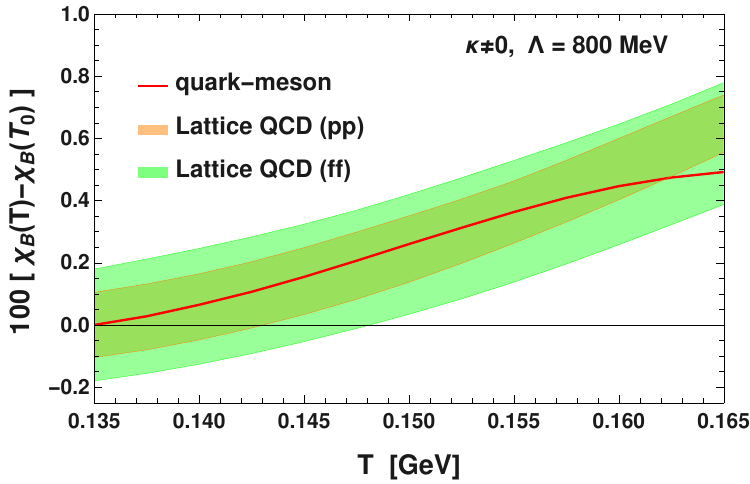}
    \caption{Subtracted susceptibility $\chi_B(T)-\chi_B(T_0)$ with $T_0=135$~MeV as a function of $T$. The lattice data are represented with the bands, while the red solid curve denotes the result from the quark-meson model (here independent of $m_0$).
    \label{fig:chiBsub}}
\end{figure}

In panel~(a) of Fig.~\ref{fig:chiBwithvacuum} we present the results for $\chi_B(T)-\chi_B(0)$ obtained with the described procedure, 
probing several values of $m_l(0)$. All the curves start at $T=135$~MeV, as this is the lower limit for the data of Fig.~\ref{fig:fittedQuarkMass}.
We note that for the temperature range $135 \ \text{MeV}<T< 155 \ \text{MeV}$, all the tested values of $m_l(0)$ produce a reasonable description of the lattice data within their large error-band. We choose $m_l(0) = 400$~MeV for further 
studies, since this value produces an acceptable agreement in~Fig.~\ref{fig:chiBwithvacuum}(a), 
while simultaneously is not far from estimates of the quark model in its application to the hadron structure.

In panel~(b), we compare the cases with and without the quark anomalous magnetic moments, drawing the conclusion that the 
role of the quark anomalous magnetic moment is significant, as it largely enhances the paramagnetic component (cf.~Eq.~(\ref{eq:chiBparaTsaiYildiz})). 

Finally, in panel~(c) we present the decomposition of 
$\chi_B(T)-\chi_B(0)$ into various components described in the legend. 
The thermal contribution for the quarks is strongly paramagnetic and increases monotonically with the temperature, 
while the thermal contribution of the mesons is diamagnetic and does not vary much in the 
considered range of $T$ (reflecting the mild growth of the pion mass with $T$ below $T_c$~\cite{Kamikado:2014bua}).
We stress the dominant role of the thermal part of the quarks in the monotonicity of the result.

The vacuum contribution of the quarks remains flat at low $T$, and above $T\approx155$~MeV  starts to decrease, which
follows from Eq.~(\ref{eq:chiB_vac}) with a decreasing quark mass. This fall-off is responsible for the bending down of the 
total contribution at $T \gtrsim 155$~MeV, as seen in panels~(a) and~(b). 
The small  mesonic vacuum contribution is positive and remains essentially constant for the entire temperature range with a slight increase above $T\approx165$~MeV,
where the pion mass grows.

To avoid the dependence on an unknown value of the quark mass at $T=0$, 
one can subtract $\chi_B$ at a temperature within the range of the data in Fig.~\ref{fig:fittedQuarkMass}, for 
instance at $T_0=135$~MeV. We can thus obtain $\chi_B(T)-\chi_B(T_0)$ and compare it with the similarly subtracted lattice data. 
In Fig.~\ref{fig:chiBsub}, we show the result of such a prescription. It can be seen that the quark-meson model reproduces the lattice data very reasonably in the displayed range of temperatures.

\begin{figure}
    \centering
    \includegraphics[width=0.47\textwidth]{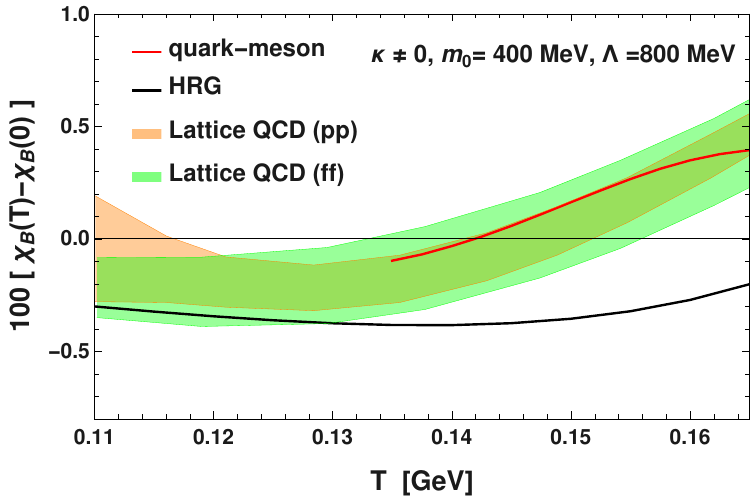}\\
    \caption{Magnetic susceptibility from the quark-meson model, from HRG, and the lattice data. The HRG result includes the anomalous magnetic moments of the hadrons. The quark-meson model (with non-zero anomalous magnetic moments of the quarks) result is with $m_0=400$ MeV and $\Lambda=800$~MeV.\vspace{1cm}\label{fig:chiBsummary}}
\end{figure}

In Fig.~\ref{fig:chiBsummary}, we summarize the results of our analysis. 
We note that the quark-meson model with suitable parameters is able to reproduce the lattice data for $\chi_B$ in the temperature range where HRG fails. 

\section{Conclusions}

Here are the main results of this paper:

\begin{itemize}

     \item Amending HRG with physical magnetic moment of hadrons does not cure its mismatch~\cite{Bali:2014kia,Bali:2020bcn} with 
     lattice data for $\chi_{B}$ at $T\gtrsim 120$~MeV.
     
     \item The contribution to $\chi_B$ from the pion--vector-meson loops, is small (5-10\%) and paramagnetic. 
     
     \item The contribution of the hadronic magnetic polarizabilities to $\chi_B$ is negligible.
   
     \item In the quark-meson approach, it is possible to describe the lattice data for $\chi_{B}$ in the vicinity of the 
     cross-over point, and {\it simultaneously}  the data  for the conserved charge susceptibilities $\chi_{\cal BB}$ and $\chi_{{\cal B}S}$.
     Our modeling incorporates the vacuum contribution from the quark loops, treated with a low-energy cut-off, and from the meson loops.
     We have accounted for the anomalous magnetic moments of the quarks, which improve the results. 
     
     \item The quark $T$-dependent masses, obtained from the precise lattice data for $\chi_{\cal BB}$ and $\chi_{{\cal B}S}$ from~\cite{Bollweg:2021vqf}, are close to
     the quasiparticle masses of Fig.~2 in~\cite{Mykhaylova:2019wci} but significantly larger (cf. Fig.~\ref{fig:fittedQuarkMass}) 
     from other quark model estimates.

\end{itemize}
   
In general, it is clear from the lattice data that to properly model $\chi_B(T)$, one needs the pions (and kaons) to 
generate diamagnetism at low $T$~\cite{Bali:2014kia,Bali:2020bcn}, and 
sufficiently light fermions (quarks), yielding enough paramagnetism to raise $\chi_B(T)$ above zero beyond the cross-over temperature. Our analysis shows 
that such a scenario is possible to realize. Moreover, we keep the agreement for $\chi_{\cal BB}$ and $\chi_{{\cal B}S}$ data from the lattice. 
More uniform quark model analysis are possible here, such as, e.g., within the Polyakov--Nambu--Jona-Lasinio model in magnetic field, 
which must be treated at the one-meson-loop 
level (i.e., involving the quark ring diagrams) to generate the physics of the pion loop. Such challenging studies, confronted to {\it all} the available lattice data, 
would be very interesting.

Also, with the deep implications of the lattice data on our understanding of the basics of hot media in the presence of a 
magnetic field, it would be very desirable to carry out future simulations leading to lower error bands in $\chi_B$.

A final remark concerns a somewhat different  topic of modeling 
the multiplicities in the phenomenology of the ultrarelativistic heavy ion reactions, which is the key success of HRG. If the relevant degrees of freedom down to 
temperatures $\approx120$~MeV involve the quarks and not the baryons,  as suggested by the lattice results for $\chi_B$, then the 
successful application of HRG for the multiplicities should involve quark fragmentation into hadrons. Such 
relevant issues are left for future considerations.

\section*{Acknowledgments}
The authors acknowledge the support from the Polish National Science Center grant 2023/51/B/ST2/01625.

%\section*{Data availability statement}

\appendix

\section{$\chi_B$ in HRG \label{s:chiB_derv}}

\subsection{$\kappa=0$ case}

The expression for the pressure of Eq.~(\ref{eq:pressurefull}) at $\kappa=0$ becomes
\begin{eqnarray}
 &&\hspace{5mm} P= \eta T \frac{B|Q|}{2\pi^2} \sum_{l=0}^\infty \sum_{s_z=-s}^s \int_0^\infty dp_z  \\
 &&\log\left[1+\eta \exp(-\frac{\sqrt{m^2+p_z^2+B [|Q|(2l\!+\!1\!)-\!2Q s_z]}}{T})\right]. \nonumber  
\end{eqnarray}
Next, applying Euler-Mclaurin summation formula~\cite{pathriaStatisticalMechanics2011}, 
\begin{eqnarray}
    \sum_{l=0}^{\infty}f(l+\frac{1}{2}) \simeq \int_0^\infty f(x) dx +\frac{1}{24}f'(0)
\end{eqnarray}
and using scaled variables  $p_z = rm$, $T=tm$ and 
\mbox{$x={z^2m^2}/{2B|Q|}$}, one obtains  
\begin{eqnarray}
 &&\chi_B = \frac{\partial^2P}{\partial B^2}\bigg|_{B=0} = \frac{ Q^2}{2\pi^2 } \left(\sum_{s_z=-s}^s s_z^2\right) \times \nonumber \\
 &&\int_0^\infty dr \int_0^\infty dz \; z \frac{\left( t+S(r,z)\right) \exp(\frac{S(r,z)}{t})+\eta\; t}{t \; S(r,z)^{3}\left(\exp(\frac{S(r,z)}{t})+\eta\right)^2} \nonumber \\
 &&\hspace{-5mm}-\frac{Q^2}{24 \pi^2} \sum_{s_z=-s}^s \int_0^\infty dr \frac{1}{S(r,0)\left( \exp(\frac{S(r,0)}{t})+\eta\right)} \; ,
\end{eqnarray}
 where $S(r,z)=\sqrt{1+r^2+z^2}$. The first term is paramagnetic (positive) and the second is diamagnetic (negative). 
 Writing them as $\chi_B^{para}$ and $\chi_B^{dia}$ we obtain
\begin{eqnarray}
    &&\hspace{-5mm}\chi_B^{dia}=-\frac{Q^2(2s+1)}{24 \pi^2} \int_0^\infty dp \frac{f(E_p)}{E_p} ,
\end{eqnarray}
where $E_p=\sqrt{p^2+m^2}$ and $f(E_p)$ is the distribution function. Similarly, 
the paramagnetic term can be calculated by carrying out the $z$-integration and an integration by parts,
\begin{eqnarray}
    &&\chi_B^{para}=\frac{Q^2\;s(s+1)(2s+1)}{6\pi^2} \int_0^\infty dp \; \frac{f(E_p)}{E_p}.
    \label{eq:chiBpara_derv}
\end{eqnarray}
Comparing with the diamagnetic part we have,
\begin{eqnarray}
    &&\frac{\chi_B^{para}}{\chi_B^{dia}} = - 4 s (s+1),
\end{eqnarray}
which for spin-$\tfrac{1}{2}$ states yields $-3$. 

\subsection{$\kappa\neq0$ case}

For $\kappa\neq0$ case, the diamagnetic contribution remains of course the same. 
The paramagnetic part depends on the spin of the particle. Let us first consider the
Tsai-Yildiz formula of Eq.~(\ref{eq:TsaiYldiz}) for spin-$\tfrac{1}{2}$ particles.
Carrying out similar steps as in the $\kappa=0$ case, we find the term linear in $\kappa$,
\begin{eqnarray}
 &&\hspace{5mm}\kappa \frac{Q s(s+1)(2s+1)}{3\pi^2} \int dz \int dr  \nonumber\\
 &&\frac{ z \left(  e^{\frac{S(r,z)}{t}}  \left( -r^2 t + S(0,z)^2 S(r,z) \right)-  r^2 t \eta\right)}{ t S(0,z) S(r,z)^{3} \left( e^{\frac{S(r,z)}{t}} + \eta \right)^2 },
\end{eqnarray}
and quadratic in $\kappa$, 
\begin{eqnarray}
&&\hspace{10 mm}\kappa^2\frac{s(s+1)(2s+1)}{6 \pi^2}  \int dr \int dz  \\
&&\frac{z \left( e^{\frac{S(r,z)}{t}} S(0,z)^4 S(r,z) - r^2 t \; S(0,z)^2 \left( e^{\frac{S(r,z)}{t}} + \eta \right)\right)}{ t S(0,z)^2 S(r,z)^3 \left( e^{\frac{S(r,z)}{t}} + \eta \right)^2}.\nonumber
\end{eqnarray}
which  after carrying out the $z$-integration and integration by parts over $r$ lead to Eq.~(\ref{eq:chiBparaTsaiYildiz}). 

For spin-1 or $\tfrac{3}{2}$ particles, with Eq.~(\ref{eq:enDelta}) the paramagnetic contribution is given by
\begin{eqnarray}
    &&\chi_{B;s=1,\frac{3}{2}}^{para} = \frac{s(s+1)(2s+1)}{6\pi^2} \int_0^\infty dp \times  \\ 
    &&~~\hspace{-2mm}\left[Q^2+2 Q \;  \left(1+\frac{p^2}{m^2}\right)^{1/2} \kappa  + \left( 1+2 \frac{p^2}{m^2}\right) \; \kappa^2 \right] \; \frac{f(E_p)}{E_p}. \nonumber 
    \label{eq:chiBparaHigherSpin}
\end{eqnarray}
Similarly, using the expression for higher spin particles of Eq.~(\ref{eq:enHighspin}) one arrives at
\begin{eqnarray}
    &&\hspace{-7mm}\chi_{B; s>\frac{3}{2}}^{para} = \frac{s(s+1)(2s+1)\left(Q+\kappa\right)^2 }{6\pi^2} \times \\ 
    &&\int_0^\infty dp  \; \left( 1+2 \frac{p^2}{m^2}\right) \; \frac{f(E_p)}{E_p}. \nonumber 
    \label{eq:chiBparaHigherSpin2}
\end{eqnarray}

\section{Self-consistency in quark models\label{s:selfcons}} 
 
In the independent particle picture, 
the decomposition of the pressure  into the thermal and vacuum parts involves functions of the following 
arguments:
\begin{eqnarray}
P=P^{\rm th}(T,\mu,B; \{m\})+P^{\rm vac}(B; \{m\}), 
\end{eqnarray} 
where $\{m\}$  denotes the collection of quark masses. 
Since the energy spectrum of particles and antiparticles is the same due to the charge conjugation symmetry, 
we have the feature
\begin{eqnarray}
P^{\rm th}(T,\mu,B; \{m\})=P^{\rm th}(T,-\mu,B; \{m\}), \label{eq:syme}
\end{eqnarray}
i.e., the symmetry in $\mu$.
Note that the vacuum term does not explicitly depend on the thermal parameters. 

The quark masses may depend on the thermal parameters and $B$, namely $\{m\}= \{m(T,\mu,B)\}$, bringing additional implicit dependence in $P$.
In effective models, $P$ is treated as an effective potential for the determination of the quark masses (or other 
effective parameters, such as the value of the Polyakov loop). 
The self-consistency condition is 
\begin{eqnarray}
\frac{\partial P}{\partial m}=0. \label{eq:self}
\end{eqnarray}
Note that thus obtained mass must be a symmetric function of $\mu$, as implied by the property (\ref{eq:syme}). In particular, 
\begin{eqnarray}
\partial m/\partial \mu \mid_{\mu=0}=0. \label{eq:mid}
\end{eqnarray}

It is now instructive to evaluate various quantities needed in this work. 
The baryon density is 
\begin{eqnarray}
{\cal B}=\frac{dP}{d\mu_{\cal B}} =\frac{\partial P}{\partial \mu_{\cal B}}+\frac{\partial P}{\partial m}\frac{\partial m}{\partial \mu_{\cal B}} = 
\frac{\partial P^{\rm th}}{\partial \mu_{\cal B}}, \label{eq:selfB}
\end{eqnarray}
where (\ref{eq:self}) was used. For the susceptibility,
\begin{eqnarray}
&& \hspace{-9mm} \chi_{\cal BB}=\frac{1}{T^2} \left . \frac{d^2P}{d\mu_{\cal B}^2}\right |_{\mu_{\cal B}=0} =
 \frac{1}{T^2}  \left [ \frac{\partial ^2P}{\partial \mu_{\cal B}^2} + \frac{\partial P}{\partial m}\frac{\partial^2 m}{\partial \mu_{\cal B}^2} \right . \nonumber \\
 && \left . + 2 \frac{\partial^2 P}{\partial m \partial \mu_{\cal B}}\frac{\partial m}{\partial \mu_{\cal B}} 
 +  \frac{\partial^2 P}{\partial m^2}\left(\frac{\partial m}{\partial \mu_{\cal B}} \right )^2 \right ]_{\mu_{\cal B}=0}  \nonumber \\
 && =\frac{1}{T^2} \left . \frac{\partial^2 P^{\rm th}}{\partial \mu_{\cal B}^2}\right |_{\mu_{\cal B}=0}, \label{eq:selfsus}
\end{eqnarray} 
where we have used conditions (\ref{eq:self}) and (\ref{eq:mid}). An analogous equality follows for $ \chi_{{\cal B}S}$:
\begin{eqnarray}
 \chi_{{\cal B}S}  =\frac{1}{T^2} \left . \frac{\partial^2 P^{\rm th}}{\partial \mu_{\cal B}\partial \mu_{S}}\right |_{\mu_{\cal B,S}=0}. \label{eq:selfsusBS}
\end{eqnarray} 
The meaning of Eqs.~(\ref{eq:selfB}-\ref{eq:selfsusBS}), despite the simplicity of the derivation, is sound, as (in the independent particle approach) 
it allows one for a ``naive'' evaluation of these quantities just from the thermal part. The vacuum term manifests itself only indirectly through the self-consistency 
conditions, in general leading to $T$, $\mu$, or $B$-dependent quark masses. 

For the magnetization and the magnetic susceptibility, derived using Eqs.~(\ref{eq:defmagnetization}) and (\ref{eq:defchiB}), 
the situation is entirely different, since both the thermal and the vacuum parts depend 
explicitly on the magnetic field.\footnote{The explicit dependence of $P^{\rm th}$ and $P^{\rm vac}$ on $B$ originates 
simply from the explicit dependence of the energy spectrum on $B$, as 
discussed in detail in Sec.~\ref{s:HRG_basics}.} Thus, the vacuum
contributes directly to ${\cal M}$ and $\chi_B$ and needs to be considered~\cite{Endrodi:2013cs,Kamikado:2014bua}. 
This makes the modeling of these magnetic features much more demanding, as it is prerequisite to model the vacuum part of $P$.

\section{Vacuum contribution to $\chi_B$ from quark loops\label{s:vacu_quark}}

We consider non-zero anomalous magnetic moment of quarks. 
%To the best of our knowledge the $\kappa \neq 0$ case has not been presented elsewhere. 
In presence of $\kappa$, the photon vacuum polarization tensor is given by 
\begin{eqnarray}
    &&\hspace{-5mm}\Pi^{\mu \nu} = i e^2 \int \frac{d^4p}{(2\pi)^4} \Tr[ \Gamma^\mu S_F(p+q) \Gamma^\nu S_F(p)] \;  , \;
    \label{eq:poltensorfermion}
\end{eqnarray}
where $\Gamma^\mu = Q\gamma^\mu + \frac{\kappa}{2m} \sigma^{\mu\nu}q^\nu$, with $\sigma^{\mu\nu}=-\frac{i}{2}[\gamma^\mu,\gamma^\nu]$. 
The quantities $p=(p_0,\vec{p})$ and $q=(q_0,\vec{q})$ are the loop momentum and the external momentum respectively,
whereas $S_F(p)$ denotes the vacuum quark propagator 
\begin{eqnarray}
    S_F(p) =  \frac{\slashed{p} + m}{p^2-m^2+i\epsilon} .
    \label{eq:propagatorferm}
\end{eqnarray}
Using Eq.~(\ref{eq:defchiB_vacpol}) yields
\begin{eqnarray}
    \chi_B^{q,\  \rm vac} =  R_0 + R_1 \kappa + R_2 \kappa^2,
    \label{eq:vac_expand}
\end{eqnarray}
where
\begin{eqnarray}
    &&R_0 =  \frac{4}{3} Q^2\pi^2 \left[(q^2+2m^2)B_0(q^2,m^2,m^2) - 2 A_0(m^2) \right] ,\nonumber \\
    &&R_1= 4\pi^2Q \; q^2 B_0(q^2,m^2,m^2),  \\
    &&R_2 = \frac{\pi^2}{6} \frac{q^2}{M^2}\left[(q^2+8m^2)B_0(q^2,m^2,m^2) - 2 A_0(m^2)  \right]. \nonumber 
    \label{eq: kapcoeff}
\end{eqnarray}
Here $A_0$ and $B_0$ are the one-loop Passarino-Veltman (PaVe) functions~\cite{Passarino:1978jh},
 \begin{eqnarray}
     &&i\pi^2 A_0(m^2) = \int d^4p \; \frac{1}{p^2-m^2+i\epsilon}, \nonumber\\
     &&\hspace{-7mm} i\pi^2 B_0(q^2,m_1^2,m_2^2) = \nonumber\\ 
     &&\int d^4p\; \bigg[\frac{1}{(p+q)^2-m_1^2+i\epsilon} \frac{1}{p^2-m_2^2+i\epsilon}\bigg].
 \end{eqnarray}
Since these integrals are divergent, one needs to use regularization. In an effective model, regularization
has a meaning of suppressing the high-momentum contributions in the loop integration. 
Here we use Pauli-Villars regularization of~\cite{RuizArriola:2002wr,Broniowski:2021awb} using two subtractions, amounting to the replacement of
$F(\{m^2\})$ with
\begin{eqnarray}
    &&\hspace{-12mm}F^\Lambda(\{m^2\}) = F(\{m^2\})-F(\{m^2 + \Lambda^2\}) \nonumber\\
    &&+ \Lambda^2 \frac{dF(\{M^2+\Lambda^2\})}{d\Lambda^2}, \label{eq:PV}
\end{eqnarray}
where $F$ denotes a PaVe function. We explicitly obtain
\begin{eqnarray}
    &&\hspace{-7 mm}A_0 = -\frac{m^2\log(\frac{m^2}{\Lambda^2+m^2})+\Lambda^2}{16\pi^4}  \\
    &&\hspace{-7 mm}B_0 = \frac{\log(\frac{\Lambda^2}{m^2}+1)-\frac{\Lambda^2}{\Lambda^2+m^2}}{16\pi^4} + 
    \frac{\Lambda^4 q^2}{96 \pi^4m^2(\Lambda^2+m^2)^2}. \nonumber
\end{eqnarray}
After collecting the terms in Eq.~(\ref{eq:vac_expand}) we obtain Eq.~(\ref{eq:chiB_vac}). 

\begin{figure}
    \centering
    \includegraphics[width=0.47\textwidth]{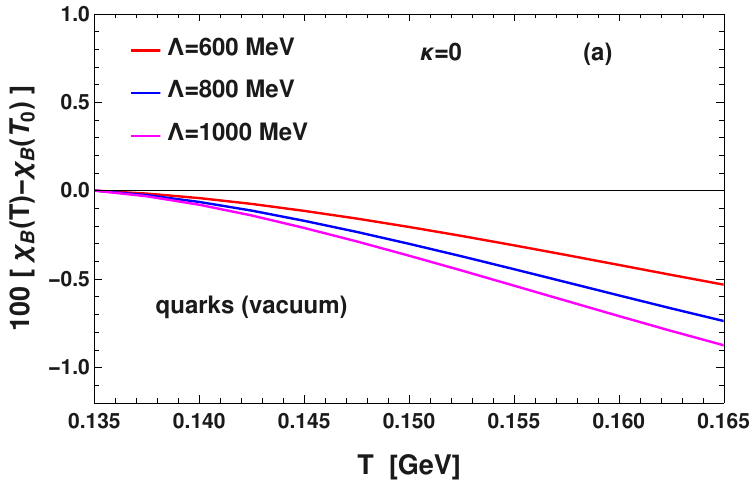}\\
    \includegraphics[width=0.47\textwidth]{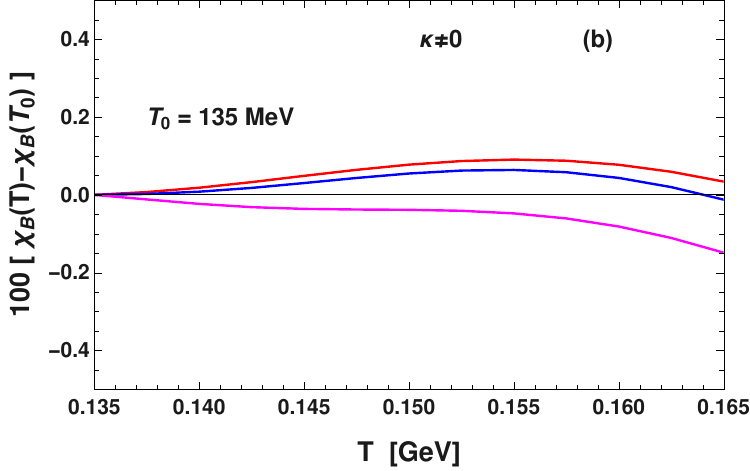}
    
    \caption{Vacuum quark contribution to $\chi_B(T)$,  subtracted at $T_0=135$~MeV, for different values of the regulator $\Lambda$, 
    with $\kappa=0$~(a) and $\kappa\neq0$~(b).
    \label{fig:vacLamda}}
\end{figure}

Lattice QCD presents the results for $\chi_B$ subtracted at $T=0$. In the absence of precise 
knowledge about $m(T=0)$ from our fit to the lattice data for $\chi_{\cal BB}$ and $\chi_{\cal BS}$, we can 
subtract the value at $T_0=135$ MeV. The results are presented in Fig.~\ref{fig:vacLamda} for different values of $\Lambda$, both for $\kappa=0$ and non-zero. 
Formally, the $\Lambda \to \infty$ limit for $\kappa=0$ reproduces the result of~\cite{Endrodi:2013cs,Kamikado:2014bua}, which may serve as a check.

\section{Pion--vector-meson loop contributions to $\chi_B$}

\subsection{$\pi-\pi$ loop \label{s:pi-pi}}

As a warm-up, let us first consider the case where the photon vacuum polarization diagram involves a pion 
loop, as drawn in Fig.~\ref{fig:pi-V loop}(a). The polarization tensor is
\begin{eqnarray}
    &&\Pi^{\mu \nu} = -i e^2 \int \frac{d^4p}{(2\pi)^4} (2p+q)^\mu  S_B(p+q) (2p+q)^\nu S_B(p), \nonumber \\
    \label{eq:poltensorpipi}
\end{eqnarray}
where the thermal propagator $S_B(p)$ is given by 
\begin{eqnarray}
    &&\hspace{-5mm}S_B(p) = \frac{1}{p^2-m^2+i\epsilon} - 2\pi i f(|p_0|)\delta(p^2-m^2).
    \label{eq:propagator}
\end{eqnarray}
Here $f$ is the Bose-Einstein distribution function of Eq.~(\ref{eq:distfunc}). 

The magnetic susceptibility is defined by Eq.~(\ref{eq:defchiB_vacpol})~\cite{Endrodi:2022wym}. After plugging in the propagator in Eq.~(\ref{eq:propagator}) inside Eq.~(\ref{eq:poltensorpipi}), taking the limit $q_0\rightarrow0$, and carrying out the sum over the spatial index, one obtains
\begin{eqnarray}
&&\lim_{q_0\rightarrow0}\Pi_s^{T\neq0} = -\pi e^2 \int \frac{d^3p}{(2\pi)^3}\frac{dp_0}{2\pi}f(|p_0|) \delta(p_0^2-E_p^2) \nonumber \\
&&\hspace{-1mm}\bigg[\frac{q^2+4p^2+4\vec{q}\cdot\vec{p}}{p_0^2-E_p^2-q^2-2\vec{p}\cdot\vec{q}+i\epsilon} 
+ \frac{q^2+4p^2-4\vec{q}\cdot\vec{p}}{p_0^2-E_p^2-q^2+2\vec{p}\cdot\vec{q}+i\epsilon}  \bigg] . \nonumber \\
\end{eqnarray}
Performing the delta function integral and taking only the positive-energy contribution one gets
\begin{eqnarray}
    &&\lim_{q_0\rightarrow0}\Pi_s^{T\neq0} = -\frac{e^2}{16\pi^2} \int p^2dp \frac{f(E_p)}{E_p} \nonumber \\ 
    &&\int_{-1}^{+1} dz \bigg[\frac{q^2+4p^2+4qpz}{-q^2-2qpz+i\epsilon} 
+ \frac{q^2+4p^2-4qpz}{-q^2+2qpz+i\epsilon}  \bigg] .
\end{eqnarray}
Keeping the real part of the above equation, carrying out the $z$-integration, and taking the limit $\epsilon\rightarrow0$ yields
\begin{eqnarray}
    &&\hspace{7mm} \lim_{q_0\rightarrow0}\text{Re}[\Pi_s^{T\neq0}] = \frac{e^2}{16\pi^2} \int p^2 dp \frac{f(E_p)}{E_p} \times \nonumber \\
    &&\bigg[ 8+ \frac{q^2-4p^2}{q\;p}\log\big[\frac{q-2p}{q+2p} \big]\bigg] .
\end{eqnarray}
Finally, by applying Eq.~(\ref{eq:defchiB_vacpol_thermal}) one arrives at Eq.~(\ref{eq:chiBdia}).

Following similar methods for the fermions, one obtains Eq.~(\ref{eq:largem}).

\subsection{$\chi_B$ from $\pi$--vector-meson loop\label{s:pi-rho}}

Similarly as in the previous subsection, one can calculate the vacuum polarization diagram of 
the pion--vector meson ($\pi-V$) loop, as drawn in Fig.~\ref{fig:pi-V loop}(b,c). The only differences are the
couplings at the vertices and unequal masses in the propagators, $m_\pi$ and $m_V$. The polarization tensor in this case is given by
\begin{eqnarray}
    &&\Pi^{\mu \nu} = -i e^2 \bigg(\frac{g^2}{m_V^2}\bigg) \int \frac{d^4p}{(2\pi)^4} G^{\mu\nu}(p,q)  S_B^\pi(p+q) S_B^V(p), \nonumber  \\
    \label{eq:poltensorpirho}
\end{eqnarray}
where   
\begin{eqnarray}
    &&\hspace{-12mm}G^{\mu\nu}(p,q) = \varepsilon^{p\alpha q \mu} \big( \frac{p^\alpha p^\beta}{m_V^2}-g^{\alpha \beta}\big) \varepsilon^{p\beta q\nu}
    \label{eq:couplingtensor}
\end{eqnarray}
and $\varepsilon^{p \alpha q \mu}=\varepsilon_{\gamma\alpha\delta\mu}p^\gamma q^\delta$. This yields
\begin{eqnarray}
    &&\Pi_s^{T\neq0} = -\pi e^2 \left( \frac{g^2}{m_V^2}\right) \int \frac{d^4p}{(2\pi)^4} \times \nonumber \\
    &&\Bigg[ \sum_{i=1}^3 G^{ii}(p,q)f(|p_0|)\frac{\delta(p^2-m_V^2)}{(p+q)^2-m_\pi^2+i\epsilon}\nonumber \\
     &&+ \sum_{i=1}^3 G^{ii}(p-q,q) f(|p_0|)\frac{\delta(p^2-m_\pi^2)}{(p-q)^2-m_V^2+i\epsilon} \Bigg]
\end{eqnarray}
and
\begin{eqnarray}
&&\hspace{-3 mm}\lim_{q_0\rightarrow0}\text{Re}[\Pi_s^{T\neq0}] = -\frac{e^2}{16\pi^2} \left( \frac{g^2}{m_V^2}\right)\int p^2dp \int_{-1}^{+1} dz \times \nonumber \\ 
&& \text{Re}\bigg[\frac{f(E_V)}{E_V}\frac{q^2(p^2z^2-p^2+2E_V^2)}{\Delta-q^2-2qpz+i\epsilon} \nonumber \\
&&\hspace{15 mm}+ \frac{f(E_\pi)}{E_\pi}\frac{q^2(p^2z^2-p^2+2E_\pi^2)}{-\Delta-q^2+2qpz+i\epsilon}  \bigg] \,
\end{eqnarray}
where $\Delta = m_V^2-m_\pi^2$.
Following similar steps as in the previous subsection yields Eq.~(\ref{eq:chiBpirho}).

\bibliography{ref}

\end{document}